# Structural neuroimaging as clinical predictor: a review of machine learning applications

José María Mateos-Pérez[*], Mahsa Dadar[*], María Lacalle-Aurioles, Yasser Iturria-Medina, Yashar Zeighami, Alan C. Evans

* These authors contributed equally to this work.

Montreal Neurological Institute, McGill University, Montreal, Quebec, Canada

## Abstract

In this paper, we provide an extensive overview of machine learning techniques applied to structural magnetic resonance imaging (MRI) data to obtain clinical classifiers. We specifically address practical problems commonly encountered in the literature, with the aim of helping researchers improve the application of these techniques in future works. Additionally, we survey how these algorithms are applied to a wide range of diseases and disorders (e.g. Alzheimer's disease (AD), Parkinson's disease (PD), autism, multiple sclerosis, traumatic brain injury, etc.) in order to provide a comprehensive view of the state of the art in different fields.

# 1. Introduction

Machine learning (ML) algorithms (Kotsiantis et al., 2007, 2006) are currently employed in an extensive range of fields, from e-mail filtering (Guzella and Caminhas, 2009), movie recommendations (Park et al., 2012) and energy grid maintenance (Rudin et al., 2012), to cite a few. In general, supervised ML consists of algorithms capable of generalizing rules or patterns from a labeled set of input data, and using that knowledge to generate predictions or classifications on data not seen before (Kotsiantis et al., 2007). The field of neuroscience has also greatly benefited from ML. For years, ML algorithms have been widely used to build classifiers or predictors for a wide range of diseases using magnetic resonance imaging (MRI) information as input features. These inputs can be structural gray matter (GM) readings, obtained from cortical thickness (CT) (Ad-Dab'bagh et al., 2006; Fischl and Dale, 2000) or GM density (GMd) values from voxel-based morphometry (VBM) (Ashburner and Friston, 2000), microstructural changes in the white matter (WM) from diffusion-weighted imaging (DWI) (fractional anisotropy (FA)) (Mandl et al., 2008), connectivity matrices (Iturria-Medina et al., 2007), or parameters derived from network analyses (Iturria-Medina, 2013; Rubinov and Sporns, 2010; Zeighami et al., 2015), and resting/task state fMRI information (Pereira et al., 2009). These values can be obtained per voxel or averaged over anatomical regions using atlases to reduce feature dimensionality. Once the imaging features have been computed, they are fed into the ML algorithm of choice in order to learn disease patterns.

Here, we present a review of publications that use structural MRI data, including DWI techniques, to build classifiers aimed both at a) predicting a given clinical state and b) extracting brain regions related to the disease of interest. As certain generalizations can be made across modalities, in some cases we refer to fMRI studies, though they will not be the main subject of this work. Readers interested in the intersection between ML and fMRI should refer to (Haynes, 2015; Pereira et al., 2009; Schrouff, 2013). While other modalities (PET, EEG, MEG) can also be used either in isolation or in conjunction with MRI data, we only focus on structural MRI, as it already offers considerable morphological findings.

While there are many studies devoted to finding group level differences, they do not necessarily imply accurate predictions and may not be very informative when it comes to predicting the clinical outcome of individual subjects (Davatzikos, 2004; Iturria-Medina, 2013; Lo et al., 2015). Furthermore, the clinical utility of imaging metrics should be assessed by their predictive power on new data samples (Gabrieli et al., 2015; Libero et al., 2015). As we want to center this review on studies that provide predictive classification, we do not include papers that only provide correlational analyses. Following the three different definitions of the term *prediction* detailed on Gabrieli et al. (2015) (section *Analytic Approaches: From Correlation to Individualized Prediction*) (Gabrieli et al., 2015), we focus on the third, in which the goodness of the method is tested on out-of-sample predictions (i.e. data that has not been used for training the model). This definition also includes cross-validation techniques, where the reported accuracy rates are more likely to generalize to out-of-sample data. In addition, this review focuses on ML techniques that work with relatively small feature sets (compared to the number of image voxels) which require feature extraction. We acknowledge that there are ML approaches that do not necessarily need this feature extraction step such as deep learning classifiers (Deng et al., 2014; LeCun et al., 2015) in which both feature extraction and classifier learning are incorporated into a unified framework (Betechuoh et al., 2006; F. Li et al., 2014; Liu et al., 2014; Payan and Montana, 2015; Suk et al., 2014, 2015; Suk and Shen, 2013; Vincent et al., 2008). However, such techniques generally require much larger datasets and more computational power, and present interpretability challenges such that they are typically regarded as *black boxes*, and for these reasons, won't be included in this review.

On a last note for the introduction, we would like to warn that it is outside the scope of this paper to provide a detailed explanation of different ML algorithms. Support vector machines (SVMs) and linear discriminants have been explained in detail in existing reviews (Lemm et al., 2011; Pereira et al., 2009). For other algorithms such as logistic regression or random forests, and for ML techniques in general, refer to Hastie et al. (Hastie et al., 2009) (https://web.stanford.edu/~hastie/ElemStatLearn/). A more introductory version of that text (James et al., 2013) is also available at http://www-bcf.usc.edu/~gareth/ISL/.

## 2. From imaging to prediction: an overview

This section provides a brief summary of the steps involved in the development of a predictive ML model using raw imaging data as input features.

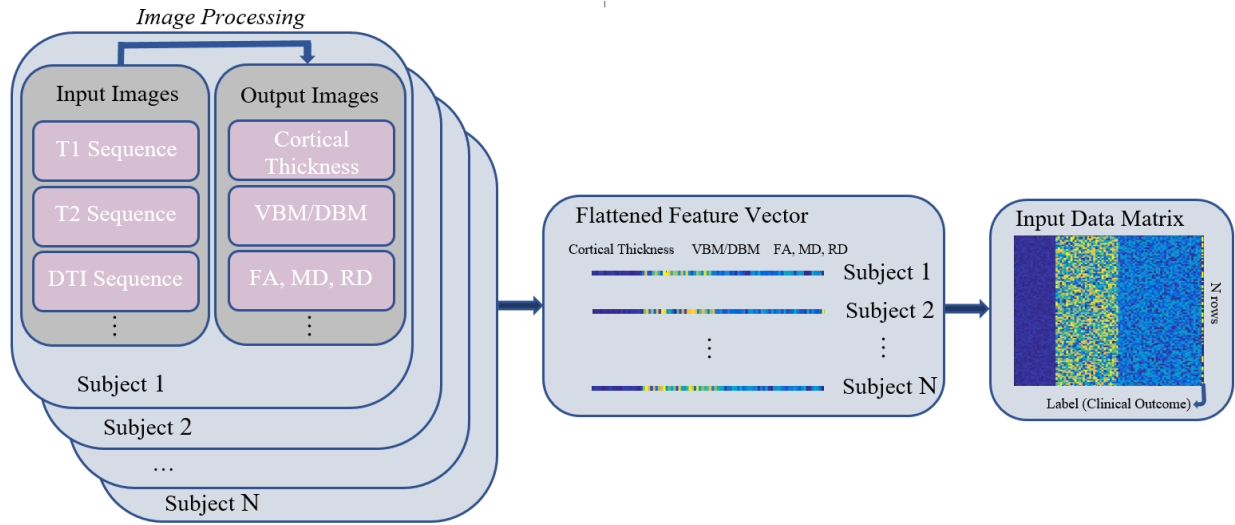

Figure 1. Image processing workflow, from the raw datasets to final input matrix for the ML system. This example assumes two different MRI modalities are used: structural T1 and DTI. The complete pipeline, from image to data input matrix, involves 3 steps: a) image processing to obtain quantitative information (e.g. CT surfaces, FA volumes, or connectivity matrices); b) removal of spatial information (*flattening*) to obtain single feature vectors per subject; and c) aggregation of all feature vectors into a single data matrix. A corresponding label output vector contains the classification target (e.g. the clinical state) for each subject. This process can involve more modalities such as PET, CSF, rs-fMRI, EEG, genetic, and behavioral information, but the final aggregated product would be similar.

## 2.1.    Image processing

Data coming from imaging studies needs be processed in order to be used as input for ML systems. This step, here referred to as *feature extraction*, typically takes place in three steps (see Figure 1 for a schematic diagram):

1. Raw images are processed to extract quantitative information. Structural T1 images can be used as the input for CIVET (Ad-Dab'bagh et al., 2006), FreeSurfer (Fischl, 2012), MINC (Aubert-Broche et al., 2013), or SPM (Penny et al., 2011) software packages, in order to extract CT per surface vertex (CIVET and FreeSurfer) or GMd per image voxel (SPM). Such processing steps generally include denoising (Manjón et al., 2010; Power et al., 2014; Wink and Roerdink, 2004), intensity inhomogeneity correction (Sled et al., 1998; Tustison et al., 2010; Vovk et al., 2007), and image intensity normalization. The images are then registered to an average brain atlas (e.g. MNI-ICBM152) (Fonov et al., 2011, 2009). Tissue or structure segmentation or cortical surface extractions are then performed using these preprocessed and normalized images in the standardized space. DWI sequences (Iturria-Medina et al., 2007; Smith et al., 2004) can also be processed using available toolboxes to extract measurements of WM microstructural changes, such as FA, mean diffusivity, radial diffusivity, connectivity matrices, and network metrics (Bullmore and Sporns, 2012). In this step, a registration procedure is also typically performed. This registration involves obtaining a series of mathematical mappings to transform the images into the same spatial domain. In other words, regardless of individual morphological differences, registration ensures that region $R$ for a given subject corresponds to the same voxels or vertices (i.e. same spatial locations) as region $R$ for the rest of the population (Hill et al., 2001; Maintz and Viergever, 1998).

2. The computed results (e.g. 3D volumetric matrices, 2D connectivity matrices, 1D vectors of network metrics, etc.) are then *flattened* in order to obtain a single feature vector per subject by removing spatial information ($x, y, z$ locations per data point) and extracting the numerical values. For instance, if CT values are computed for 40000 vertex points, a 40000×1 vector is generated, regardless of the position of the vertices within the computed surface. The necessary information to revert the values to their original spatial locations can be stored.

3. Feature vectors from all subjects are then aggregated into a $N \times M$ matrix, where $N$ is the number of subjects in the study and $M$ is the length of the feature vector, which can also include information from sources other than imaging (demographics, behavioral, etc.). Finally, the output label containing the clinical states of the subjects to be used as the target variable.

## 2.2. Building a predictive model

This subsection provides a summary on how to apply ML algorithms to processed data, such as the data matrix obtained in the previous step, as well as brief comments on potential pitfalls/aspects that might prove useful in practice. For more extensive reviews, see (Lemm et al., 2011). A ML classification algorithm is an *a priori* unknown function that relates a set of inputs with an output label (in this case, the clinical status of the subjects that form the sample). That function is then *trained* on a set of known data to obtain the parameters that relate the input vectors to categorical output values, therefore producing a classification output. This process is not sufficient by itself, as the classifier needs to be tested in a dataset not used during the training phase. Since imaging data is generally scarce, it is not common to have testing data reserved. Instead, cross-validation (Duda et al., 2001; Hastie et al., 2009) is typically used: the full dataset is split into $N$ different folds: $N-1$ are assigned for training and the remaining one for testing. The algorithm is trained and an accuracy score (e.g. percent of correctly classified subjects, sensitivity, specificity or other suitable metrics) is reported on the test set. The process is repeated until each fold has been assigned once to the test set to obtain an overall accuracy score. If the number of folds is the equal to the number of subjects in the sample, this process is called leave-one-out cross-validation, as each subject is tested individually.

## 2.3. Model ensembling and stacking

Ensembling and stacking techniques allow to combine different models (and even several instances of the same model, with different initialization parameters) in order to achieve higher accuracies and, at the same time, reduce the probability of overfitting (Hastie et al., 2009). Ensembling refers to the combination of predictions by (weighted) averaging their results, or using a voting schema (Caruana et al., 2004). On the other hand, model stacking uses the output from different classifiers as the input of another algorithm which yields the final classification score (Džeroski and Ženko, 2004). This last algorithm can be any of the ML algorithms whose results are being merged, or a completely different one.

While these approaches may yield better and more robust results than any of the best models individually, it has to be taken into account that the interpretability of the resulting classifier might not be as straightforward as it would normally be with a single model. As mentioned previously, in this field, accuracy rates are important, but so is the interpretability of the biological causes of the different diseases or disorders, such as which regions are particularly relevant for a given classification task. As a result, one might opt for a single model with slightly lower accuracy in favor of higher interpretability. In some cases, an ensemble approach may also provide feature importance as the output. For instance, random forests are by themselves an ensembling approach (a combination of individual decision trees).

# 3. Practical issues

Missteps in performing cross-validation commonly lead to overly optimistic error rates (i.e. the classifier is reported to do better than it actually does). Thus, this step should be implemented with extensive care. In the following section, we comment on details that need to be taken into consideration when implementing cross-validation loops in ML pipelines. The optimal workflow for building a robust ML classifier is depicted in Figure 2 (Gabrieli et al., 2015). For more information, see Appendix A from (Plitt et al., 2015).

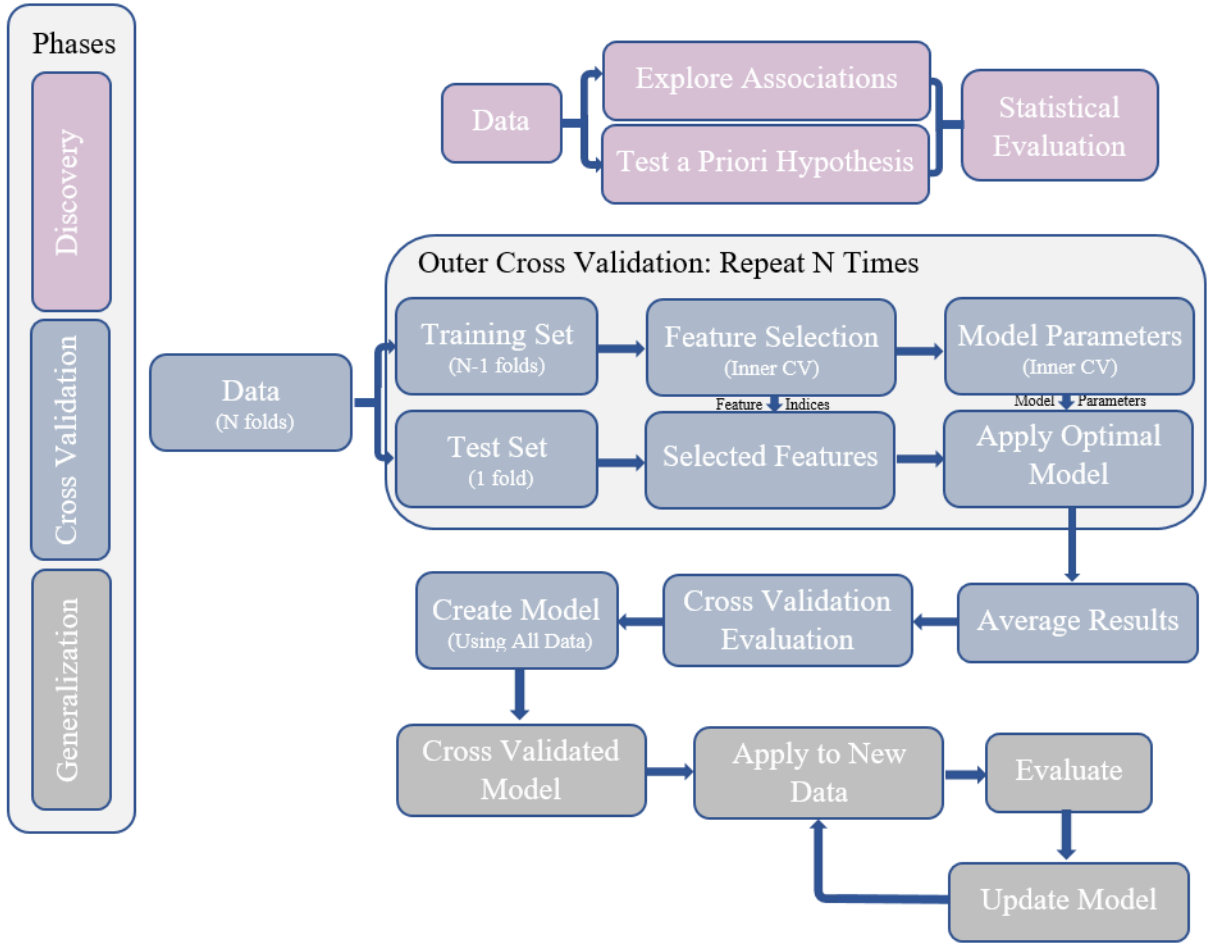

Figure 2: Optimal workflow for constructing a classifier or predictor. Splitting the data into N folds using a cross-validation approach is not the only step required to ensure generalizability. Internal cross-validation loops are necessary to obtain a subset of relevant features (if feature selection is needed) and to tune model hyperparameters (e.g. *C* in Gaussian SVMs, number of neurons in neural networks, or number of trees in random forests). Performing these steps on the full sample will result in an excessively optimistic classifier. Additionally, the cross-validation evaluation could be enhanced by performing permutation tests (Golland and Fischl, 2003; Ojala and Garriga, 2010). Figure reproduced with permission from the original (Gabrieli et al., 2015).

## 3.1. Feature preprocessing

Once the data matrix has been formed, it can be beneficial to perform an initial feature preprocessing before proceeding with the main ML pipeline. As certain algorithms expect the features to represent data in the same scale and with a certain distribution, it is common to perform a centering and scaling operation: each continuous variable is replaced by new values, obtained from subtracting the original mean and dividing by the original standard deviation (i.e. creating variables with mean=0 and standard deviation=1). During this phase, dimensionality reduction algorithms can be used, such as principal component analysis (PCA) or independent component analysis (ICA) (Duda et al., 2001; Hastie et al., 2009). While ICA is frequently used in fMRI data analysis, few studies use these techniques in the literature included in this review. This may be due to the fact that PCA and similar methods yield new variables which are linear combinations of the original ones, and hence come at the cost of reduced interpretability of the features. Other more complex feature selection techniques such as sparse feature selection can also be used in this step, depending on the specific application and dataset (Ahsen et al., 2017; Z. Li et al., 2014; Tan et al., 2010).

Depending on the application, more specific preprocessing steps may be performed, specially when a large confounding effect is encountered. Building classifiers to differentiate AD versus healthy controls, Dukart et al. found that misclassified patients were younger than misclassified control subjects (Dukart et al., 2011). Removing age-related effects from the input VBM data improved accuracy by approximately 2%. A slightly larger effect (5%) was later observed using the same technique applied to mild cognitive impairment (MCI) subjects when predicting their conversion status to AD (Moradi et al., 2015).

## 3.2. Feature selection and hyperparameter tuning

The result of the image processing step typically consists of data matrices of relatively small numbers of rows (corresponding to subjects) with significantly larger numbers of columns (corresponding to different variables), sometimes several orders of magnitude higher (e.g. several hundreds of subjects, at best, and thousands or tens of thousands of variables). These variables can be CT, GMd, or VBM measures for each voxel, or FA values in the WM. For instance, CT values extracted using the CIVET software (Ad-Dab'bagh et al., 2006) consist of more than 160,000 vertices per subject if high-resolution surfaces are used. In order to initially reduce the number of features, from thousands to just a few hundreds, it is common to use ROI-based approaches: voxels or surfaces are averaged over regions defined by a brain atlas, such as AAL (Tzourio-Mazoyer et al., 2002) or DKT (Klein and Tourville, 2012). Note that this averaging might result in losing potential differences in cases where the defined regions are too large (Dyrba et al., 2015).

As it is known, not all diseases affect every brain region, and not always in the same way. Therefore, some of the input variables might not be related to the output labels and some of them may contain information already conveyed by other features. Reducing the number of both irrelevant and redundant variables reduces the computational time improves generalization (Dash and Liu, 1997; Guyon and Elisseeff, 2003; Moradi et al., 2015). In the field of neuroscience, feature selection is relevant not only because it helps to achieve higher accuracy rates (Ad-Dab'bagh et al., 2006), but also, and mainly, because it allows to investigate which features are relevant for the specific classification problem of interest, offering an insight to the underlying brain regions that account for group differences (Plitt et al., 2015). This interpretation can make ML results complementary to those obtained by more classical inferential approaches. From this point of view, it is also important to note that some ML algorithms (e.g. linear SVMs and random forests) assign to each variable, a weight which is directly related to their importance within the model. Said weights can then be used to rank the input variables and create maps of brain regions relevant for the classification task, even when no feature selection is performed *a priori*. Storing the spatial information for the features, it is possible to report this feature importance using a parametric map (Figure 3). In that sense, certain ML algorithms can also be used as feature selection methods (Rakotomamonjy, 2003) in combination with techniques such as Recursive Feature Extraction (RFE) (Kuncheva and Rodríguez, 2010). While SVMs are capable of dealing with multiple irrelevant features (Lemm et al., 2011; Zarogianni et al., 2013), their accuracy is nonetheless diminished compared to an optimal situation in which only relevant features are used (Liu et al., 2012).

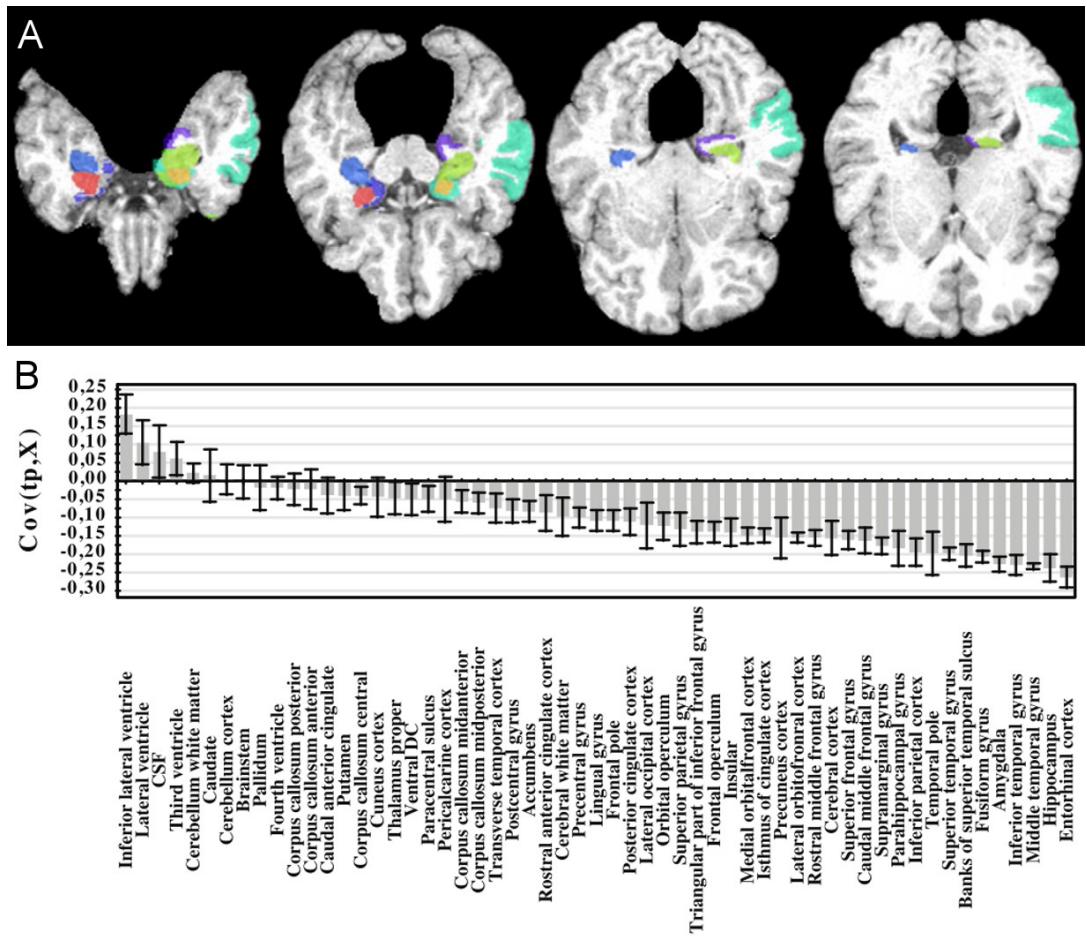

Figure 3: **A)** Using a SVM multi-kernel approach, Zhang et al. (Zhang et al., 2011) found 11 relevant cortical regions for AD classification: left and right amygdala, left and right hippocampal formations, left and right uncus, left entorhinal cortex, left middle temporal gyrus, left temporal lobe, left perirhinal cortex and left parahippocampal gyrus. This assessment of the importance of different features supports the usage of ML techniques in order to understand the biological bases of diseases. **B)** It is also possible to report variable importance without using the spatial distribution. Figure reproduced with permission from the original (Westman et al., 2012; Zhang et al., 2011).

Within the same family of classifiers, the number of features used may also have an impact. Song et al. found that Gaussian SVMs behaved better than linear SVMs in lower dimensionality problems (fewer features) (Song et al., 2011). Non-linear SVMs can be more prone to overfitting (finding noisy patterns that do not improve generalization) (Wottschel et al., 2015). In such cases, they may behave better (i.e. higher validation accuracy) if the dimensionality of the problem is reduced.

### 3.2.1. Leakage in cross-validation techniques

Leakage (Johnston et al., 2013; Kuncheva and Rodríguez, 2010; Pereira et al., 2009) is the creation and usage, commonly by accident, of variables that carry information about the outcome of the problem (the classification labels, in our case). Leakage generally occurs during feature selection if the entire dataset is used to identify potentially informative variables outside of the cross-validation loop.

A rule of thumb can be established to detect leakage. Consider the case of leave-one-out cross-validation, in which the $i^{th}$ case, denoted $X_i$, (intuitively, the $i^{th}$ row in the input matrix from Figure 1) is kept as the test set and the rest is used as the training set for that case. In that schema, involving the label for the test case ($y_i$) in any step during training would be leakage; $y_i$ should only be used when evaluating the accuracy of the model (Lemm et al., 2011). This includes somewhat common procedures such as performing t-tests, correlations or more advanced feature selection techniques on the entire sample in order to identify features strongly related to the output label before proceeding with the cross-validation. Selecting variables with the highest variance from the whole sample, on the other hand, would not be considered as leakage since output labels are not used. Sections 7.10 of Hastie et al. (*Cross-Validation*) and 7.10.2 (*The Wrong and Right Way to Do Cross-validation*) (Hastie et al., 2009) provide an overview on cross-validation. (Ambroise and McLachlan, 2002) also provides extensive comments on feature selection for microarray gene-expression data, a quintessential example in which the number of samples is much lower than the number of features. This dimensionality problem is not different from the one encountered in neuroimaging field where similar precautions may be applied.

It is not rare to detect leakage, as we will explore in the following sections. While this does not invalidate the reported findings, it makes comparison of the results difficult, as the reported accuracies will likely be overly optimistic.

Similar considerations should also be applied for hyperparameter tuning (i.e. the inherent parameters of ML algorithms, such as C in Gaussian SVMs, number of neurons in neural networks, or number of trees in random forests). To avoid synthetically increasing the accuracy, this procedure also has to be done in an inner loop within each cross-validation fold, or the model selection would be done based on the entire sample. As in the case of feature selection, it is also possible to report summary statistics about optimal hyperparameter values (Cuingnet et al., 2011).

### 3.2.2. Bias-Variance Trade-off

The relationship between the sample size (i.e. the number of subjects) and the dimensionality of the problem (i.e. the number of features) has been extensively studied in the literature (Hastie et al., 2009; Hughes, 1968; Kanal and Chandrasekaran, 1971; McKnight et al., 2002). As was mentioned previously, the number of features that are extracted from MR images are generally much larger than the sample size. In such high-dimensional cases, if the model parameters are estimated to fit the data without any form of regularization (e.g. PCA), there will be a high likelihood of overfitting to the training data, and consequently a poor generalization to out-of-sample test data (Hastie et al., 2009). On the other hand, too much regularization (e.g. using a very small number of features) might also lead to underfitting; not using all the available information from data. Determining the optimal amount of regularization is a bias-variance problem based on the sample size and specific task of interest, a high variance leads to overfitting, while a high bias leads to underfitting (Raudys and Jain, 1991). For more information of model selection, see Varoquaux et al. (Varoquaux et al., 2017).

# 4. Machine learning applied to structural neuroimaging

In the following subsections, we will discuss in more depth works that report classifiers built for specific diseases or disorders. In a few cases, publications that deal with non-categorical output variables (such as ADOS scores in autism) have also been included, but they are the exception. Classification accuracy is defined as the percentage of correct predictions; i.e. the sum of true positive and true negative predictions divided by the total number of predictions. For consistency, all accuracy scores are reported as a number between 0 and 1 (e.g. 0.67) instead of a percentage.

## 4.1. Alzheimer's disease/ mild cognitive impairment

Alzheimer's disease (AD) is a progressive neurodegenerative disorder leading to mild cognitive impairment (MCI) and dementia. The increased knowledge of the clinical manifestations and the complex biology of AD has led to the redefinition of the different disease stages in 2011 (Albert et al., 2011; McKhann et al., 2011).

Although MCI was traditionally considered as a risk factor for developing AD (Boyle et al., 2006), now it has been proposed that MCI patients who progress to AD should be reclassified as prodromal AD. On the contrary, patients who do not progress to dementia and do not show common biomarkers of AD should be considered as MCI patients (Dubois et al., 2010). Here, in order to be consistent with the terminology used in the reviewed publications, we maintain the traditional terminology, referring as MCI converted (MCI-c) to those that progress to AD dementia, and MCI non-converted (MCI-nc) to those that do not. Also, only classificatory studies of sporadic AD have been reviewed.

Typically, works on this topic have been focused on three different classification targets: a) AD patients vs. healthy controls; b) AD or controls vs. patients with MCI; and c) identification of MCI patients that will progress to AD within a certain time period (MCI-nc vs. MCI-c). The first two classification tasks address disease diagnosis, whereas the third addresses prognosis of the likely course of the disease. AD vs. MCI classification is by itself a more difficult problem than AD vs. controls (see Figure 2 from (Zhang et al., 2011)), as MCI diagnosis sometimes sits in a *gray area* (Iturria-Medina, 2013) and can be easily confounded with either mild AD or healthy controls. It is even more challenging to predict which MCI patients will progress to AD within a certain time window (typically ranging between 6 months and 3 years) and which will remain stable (Wee et al., 2012; Westman et al., 2012). Table 1 provides a summary of these papers, including input data modality, the algorithm used and achieved accuracies. In the following, relevant aspects from some of the listed works are discussed.

Table 1: Summary of the classification papers in Alzheimer's disease. Unless otherwise noted, reported accuracy rates are the highest found in the paper for different groups, methods and input modalities. "*" indicates that accuracies have been computed using sensitivity and specificity values from the paper (*accuracy = sensitivity · prevalence + specificity · (1 − prevalence)*). The value for *prevalence* has been obtained from the number of cases for each group.

| Reference | Groups (N) | Method | Input Modalities | Accuracy | Comments |
|---|---|---|---|---|---|
| (Klöppel et al., 2008) | Controls-AD (several groups) | SVM (linear) | T1 | Up to 0.964 | Independent samples for training and testing |
| (Moradi et al., 2015) | MCI-c (100) - MCI-nc (164) | SVM LDS | T1, cognitive | 0.66 0.745 | Features selected on independent AD vs. Control samples. |
| (Dyrba et al., 2015) | Controls (25) – AD (28) | SVM (Gaussian) SVM (multi-kernel) | rs-fMRI DTI T1 rs-FMRI, DTI, T1 DTI, T1 | 0.74 0.85 0.81 0.79 0.85 | AUC = 0.8 AUC= 0.87 AUC = 0.6 AUC= 0.82 AUC= 0.89 |
| (Cuingnet et al., 2011) | Controls (162) – AD (137) Controls (162) – MCI-c (76) MCI-c (76) – MCI-nc (134) | Multiple classifiers tested, linear SVM had best accuracy. | T1 | 0.82* 0.76* 0.62* | MCI conversion or non-conversion at 18 months. Results for MCIc vs. MCI-nc are non-significant. |
| (Desikan et al., 2009) | Controls (143) – MCI (113) | Logistic regression | T1 | 0.823* | Independent samples for training and testing (AUC=0.95). |
| (Zhang et al., 2011) | Controls (52) – AD (51) Controls (51) – MCI (99) | SVM (linear) SVM (multikernel) SVM (linear) SVM (multikernel) | T1 PET CSF T1, PET, CSF T1 PET CSF T1, PET, CSF | 0.862 0.865 0.821 0.932 0.72 0.716 0.714 0.745 | No hyperparameter search. Fixed C = 1. |
| (Wee et al., 2011) | Controls (17) – MCI (20) | SVM (linear) SVM (Gaussian) | DTI | 0.667 0.889 | Leakage (features selected on full dataset). |
| (Schmitter et al., 2015) | Controls (229)-AD (188) Controls (229)-MCI (401) MCI (401)-AD (188) MCI-nc (130)-MCI-c (111) (2 y) MCI-nc (103)-MCI-c (137) (3 y) | SVM (linear) | T1 | 0.883* 0.779* 0.687* 0.688* 0.698* | Leakage (voxels excluded based on statistical assessment on full dataset). Data mixed 1.5T and 3T images and was processed with 3 different software packages (FreeSurfer, MorphoBox, SPM). SVM hyperparameter search performed outside cross-validation loop. |
| (Beheshti and Demirel, 2015) | Controls (130) – AD (130) | SVM (linear) SVM (Gaussian) | T1 | 0.896 0.893 | |
| (Dukart et al., 2011) | Controls (79) – AD (80) | SVM (linear) | T1 | 0.832 | Effect of age is removed from data. |
| (Gray et al., 2013) | Controls (35) – AD (37) Controls – MCI (75) MCI-nc (41) – MCI-c (34) | Random forest | T1, CSF, genetic, FDG-PET | 0.89 0.746 0.580 | |
| (Davatzikos et al., 2011) | MCI-nc (170) – MCI-c (69) | SVM (linear kernel) | T1, CSF | 0.734 | Part of the features come from a method trained on Controls vs. AD. |
| (Davatzikos et al., 2008) | Controls (15) – MCI (15) | SVM (linear kernel) | T1 | 0.9 | Longitudinal dataset. |

| Reference | Subjects | Classifier | Input | Accuracy | Notes |
|---|---|---|---|---|---|
| (Aguilar et al., 2013) | Controls (110) – AD (116) MCI-nc (98) – MCI-c (21) | OPLS SVM (Gaussian kernel) Decision trees Neural networks OPLS SVM (Gaussian kernel) Decision trees Neural networks | T1, genetic, demographic | 0.876 0.867 0.827 0.872 0.747 0.709 0.674 0.701 | Accuracies for Controls – AD for different classifiers may use different input data. |
| (Oliveira Jr et al., 2010) | Controls (20) – AD (15) | SVM (Gaussian) | T1 | 0.882 | Hyperparameters optimized in the outer loop. Features selected on full dataset. |
| (Westman et al., 2011) | Controls (112) – AD (117) Controls (112) – MCI (122) MCI (122) – AD (117) | OPLS | T1 | 0.92* 0.769* 0.71* | |
| (Westman et al., 2012) | Controls (111) – AD (96) Controls (111) – MCI (162) MCI-nc (81) – MCI-c (81) | OPLS | T1, CSF | 0.918 0.776 0.685 | |
| (Xu et al., 2015) | Controls (117) – AD (113) Controls (117) – MCI (110) MCI-nc (83) – MCI-c (27) | wmSRC | T1, PET | 0.948 0.745 0.778 | |
| (Wee et al., 2012) | Controls (17) – MCI (10) | SVM (multi-kernel) | DTI, fMRI | 0.963 | Features selected on full dataset. |
| (Nir et al., 2015) | Controls (50) – AD (37) Controls (50) – late MCI (39) | SVM (Gaussian) | DTI | 0.849 0.79 | Features selected on full dataset. |
| (Fan et al., 2008a) | Controls (66) – AD (56) Controls (66) – MCI (88) MCI (88) – AD (56) | SVM (linear) | T1 | 0.965 0.846 0.759 | |
| (Young et al., 2013) | MCI-nc (96) – MCI-c (47) | Gaussian Process | T1, PET, APOE, CSF | 0.643 | Trained on healthy subjects + AD, tested on MCI cohort. |
| (Salvatore et al., 2016) | Controls (162) – AD (137) Controls (162) – MCI-c (76) MCI-nc (134) – MCI-c (76) | SVM | T1 | 0.92 0.86 0.73 | |
| (Duchesne et al., 2009) | Controls (75) – AD (75) | SVM (linear) | T1 | 0.92 | Other classifiers used (not reported here). |
| (Sørensen et al., 2017) | Controls (282) MCI (283) AD (154) CADDementia Test (354) | LDA | T1 | 0.63 | Multi-class classification results. Winner of CADDementia challenge. More complex classifiers did not improve performance. |
| (Ahmed et al., 2017) | Controls (52) – AD (45) Controls (52) – MCI (58) MCI (58) – AD (45) | SVM (Gaussian) | T1, DTI | 0.902 0.794 0.766 | Use a multiple kernel learning method to combine features from T1, DTI, and CSF. |
| (Vemuri et al., 2008) | Controls (190) – AD (190) | SVM (linear) | T1 T1, APOE | 0.885 0.893 | Model selection and optimization performed on 280 samples and validated on the remaining 100. |

Dyrba et al. used a multimodal approach, with T1, DTI and rs-fMRI as inputs for SVM classifiers (Dyrba et al., 2015). Using only structural T1 information, an accuracy of 0.82 was obtained. The addition of DTI increased the AUC (0.89 vs. 0.86), but no improvement was observed by using all three modalities (accuracy 0.79, AUC 0.82). Authors discuss that this could be due to high levels of noise in the rs-fMRI data, which caused SVMs to overfit during training. This provides an example where more features do not necessarily imply more validation accuracy. The authors also comment on the hypothetical existence of a *ceiling effect* which makes it impossible to obtain diagnostic accuracies significantly higher than 0.90. This observation follows the same direction as the increasingly accepted idea that AD can have a combined etiology (vascular and neuronal) which increases the variability in the burden of vascular or neuronal damage in patients with identical dementia ratings. This theoretical upper limit is well in line with the values obtained for all the other papers analyzed in this review.

Klöppel et al. obtained high accuracy rates (0.811-0.964) when comparing controls to AD patients, using training and testing datasets from different databases (Klöppel et al., 2008). As mixing images from different sites can potentially have a confounding effect (Auzias et al., 2016), this implies robustness of the selected approach and that SVMs are able to generalize well. The same approach was used in Desikan et al. (Desikan et al., 2009) for controls vs. MCI classification: 49 controls and 48 MCI patients (training) were obtained from the OASIS database (Marcus et al., 2007), and 94 controls and 57 MCI patients (test) came from ADNI (http://adni.loni.usc.edu/). They also obtained high accuracy scores on the test dataset (AUC = 0.95, sensitivity = 0.73, specificity = 0.94).

Changing the SVM kernels (linear vs. Gaussian) in Klöppel et al. had no effect on the outcome (Klöppel et al., 2008), whereas in Wee et al., a linear kernel obtained significantly lower accuracy than a Gaussian kernel (0.67 vs. 0.89) (Wee et al., 2011). However, Wee et al.

(Wee et al., 2011) suffers from leakage, as they selected features based on the entire dataset, as opposed to Klöppel et al. (Klöppel et al., 2008). Similar leakage problems are also present in other studies, such as Haller et al. (Haller et al., 2010) and Plant et al. (Plant et al., 2010).

Another measure of robustness is mixing images obtained with different field strengths. Schmitter et al. mixed structural T1 images acquired at 1.5 T and 3 T and compared a wide range of conditions (controls vs. AD, controls vs. MCI, MCI vs. AD, MCI-nc vs. MCI-c at 2 years, MCI-nc vs. MCI-c at 3 years) (Schmitter et al., 2015). In line with the rest of the literature, they report the highest accuracy for controls vs. AD classification (0.883), and the lowest for MCI vs. AD (0.687). Similar accuracies were obtained for the MCI prognosis tasks (0.688 at 2 years, 0.698 at 3 years). Authors report that using 1.5 T and 3 T datasets independently yielded similar accuracy scores.

Westman et al. (not reported in Table 1) assessed whether conversion of MCI-c patients to AD could be predicted, depending on the time window. For 12, 18, 24 and 36 months, 82.9%, 86.4%, 75.4% and 68% of MCI subjects were identified as AD patients, respectively (Westman et al., 2012).

As mentioned in the Introduction, application of ML in neuroimaging involves extracting features from the raw images. Cuingnet et al. explored the changes in accuracy when using different image processing tools for a variety of binary classification problems: AD vs. controls, controls vs. MCI-c, and controls vs. MCI-nc within an 18-month time frame (Cuingnet et al., 2011). They report a sensitivity difference of up to 0.3 in some cases due exclusively to the imaging processing technique employed.

Moradi et al. report that feature selection can improve accuracy rates up to 5%. They also identify relevant features using a controls vs. AD classification task and then use those features for classifying MCI-nc vs. MCI-c, reaching accuracy scores of up to 0.745 (AUC = 0.766) (Moradi et al., 2015), effectively showing that regions affected by AD can be useful in MCI-nc vs. MCI-c classification. Similarly, Davatzikos et al. extract regions of importance from a cohort of AD and healthy controls (Fan et al., 2008b) and applies the obtained patterns to MCI-nc vs. MCI-c classification task, also obtaining high accuracy scores (ACC = 0.734) (Davatzikos et al., 2011). This overlap in regions of importance has also been reported elsewhere (Aguilar et al., 2013; Cuingnet et al., 2011; Desikan et al., 2009; Westman et al., 2012).

Sørensen won the CADD Dementia challenge by building a multi-class LDA classifier to differentiate control, MCI, and AD simultaneously. They trained the classifier using data from more than 600 subjects from a combination of different datasets to obtain a multi-class accuracy of 0.63 on the unobserved CaDD Dementia test dataset (Sørensen et al., 2017). They further report on the effect of the size of the training dataset as well as complexity of the classifier on the performance of the classifier (Sørensen et al., 2017).

In the papers reviewed here, we found few examples of stacking/ensembling techniques. For instance, Moradi et al. first created a classifier with the imaging data and then used its output, along with age and behavioral data, as inputs to a random forest (Moradi et al., 2015). Zhang et al. combined different data sources with different SVM kernels (Zhang et al., 2011). Liu et al. also used multiple *weak* classifiers and combined their answers to produce a final result (Liu et al., 2012). Ingalhalikar et al. used this technique for a different application: to cope with missing data; different classifiers were created per subject, depending on the subset of data missing, and their outputs were merged afterwards (Ingalhalikar et al., 2014).

Table 2 summarizes the relevant GM and WM regions reported for the classification tasks in the reviewed literature. Note that different studies have followed different methodologies, some of which include selecting features based on the entire dataset, therefore creating variables that are informative at the group level, but not necessarily at the individual level. Having said that, this table paints a clear picture of AD: hippocampus, temporal lobes, amygdala, parahippocampal gyrus, middle temporal gyrus, entorhinal cortex and insula are the most important GM regions for the classification task. While fewer studies have used DTI, recent findings report microstructural WM changes and impaired connectivity as key factors leading to cognitive failure in AD. Changes in FA and mean diffusivity (MD) appear early in the disease and seem to be independent of GM changes in the medial temporal lobe (Fletcher et al., 2014; Lacalle-Aurioles et al., 2016). Decreased FA and increased MD have been described in preclinical phases of AD, when individuals are still cognitively normal; however, they have not been used in ML classification tasks at these stages (Fletcher et al., 2013).

Table 2: Informative regions (GM and WM) the classification tasks in AD. This table does not make any distinction regarding the cohorts involved in the classification (AD, MCI, controls), as it has been shown that affected regions are similar for AD and MCI.

| Region | References | N |
|---|---|---|
| Grey Matter | | |
| Hippocampus | (Aguilar et al., 2013; Beheshti and Demirel, 2015; Cuingnet et al., 2011; Davatzikos et al., 2011, 2008; Desikan et al., 2009; Dukart et al., 2011; Dyrba et al., 2015; Fan et al., 2008b, 2008a; Gray et al., 2013; Moradi et al., 2015; Schmitter et al., 2015; Westman et al., 2012, 2011, 2011) | 18 |
| Temporal lobes | (Davatzikos et al., 2011; Desikan et al., 2009; Dukart et al., 2011; Fan et al., 2008b; Moradi et al., 2015; Schmitter et al., 2015; Wee et al., 2012; Westman et al., 2012, 2011; Zhang et al., 2011) | 10 |
| Amygdala | (Aguilar et al., 2013; Cuingnet et al., 2011; Davatzikos et al., 2011; Desikan et al., 2009; Dyrba et al., 2015; Gray et al., 2013; Moradi et al., 2015; Wee et al., 2012; Westman et al., 2012; Zhang et al., 2011) | 11 |
| Parahippocampal gyrus | (Aguilar et al., 2013; Cuingnet et al., 2011; Desikan et al., 2009; Klöppel et al., 2008; Moradi et al., 2015; Oliveira Jr et al., 2010; Wee et al., 2012; Westman et al., 2012; Zhang et al., 2011) | 9 |
| Middle temporal | (Aguilar et al., 2013; Cuingnet et al., 2011; Desikan et al., 2009; Dukart et al., 2011; Gray et al., 2013; Oliveira Jr et al., 2010; Westman et al., 2012; Zhang et al., 2011) | 8 |
| Entorhinal cortex | (Aguilar et al., 2013; Cuingnet et al., 2011; Davatzikos et al., 2011; Desikan et al., 2009; Fan et al., 2008b; Oliveira Jr et al., 2010; Westman et al., 2012; Zhang et al., 2011) | 8 |
| Insula | (Aguilar et al., 2013; Davatzikos et al., 2011; Fan et al., 2008b; Moradi et al., 2015; Wee et al., 2012, 2011) | 6 |
| Inferior temporal | (Cuingnet et al., 2011; Desikan et al., 2009; Dukart et al., 2011; Fan et al., 2008a, 2008b; Westman et al., 2012) | 6 |
| Posterior cingulate | (Cuingnet et al., 2011; Davatzikos et al., 2011; Dukart et al., 2011; Fan et al., 2008b, 2008a; Wee et al., 2012) | 7 |
| Frontal lobes | (Dukart et al., 2011; Moradi et al., 2015; Wee et al., 2012) | 3 |
| Inferior parietal | (Beheshti and Demirel, 2015; Cuingnet et al., 2011; Desikan et al., 2009) | 3 |
| Anterior cingulate | (Beheshti and Demirel, 2015; Dukart et al., 2011; Wee et al., 2012) | 3 |
| Supramarginal gyrus | (Cuingnet et al., 2011; Desikan et al., 2009) | 2 |
| Middle cingulate | (Cuingnet et al., 2011; Dukart et al., 2011) | 2 |
| Thalamus | (Cuingnet et al., 2011; Dukart et al., 2011; Wee et al., 2012) | 3 |
| Uncus | (Fan et al., 2008a; Zhang et al., 2011) | 2 |
| Superior frontal lobe | (Oliveira Jr et al., 2010) | 1 |
| Parietal cortex | (Klöppel et al., 2008; Sørensen et al., 2017) | 2 |
| Cerebellar areas | (Moradi et al., 2015) | 1 |
| Posterior middle frontal | (Cuingnet et al., 2011) | 1 |
| Fusiform gyrus | (Cuingnet et al., 2011) | 1 |
| Lingual | (Desikan et al., 2009) | 1 |
| Precuneus | (Davatzikos et al., 2008; Desikan et al., 2009; Wee et al., 2012, 2011) | 6 |
| Superior temporal | (Aguilar et al., 2013; Davatzikos et al., 2011; Desikan et al., 2009; Dukart et al., 2011; Fan et al., 2008a; Oliveira Jr et al., 2010) | 6 |
| Perirhinal cortex | (Zhang et al., 2011) | 1 |
| Rectus gyrus | (Wee et al., 2011) | 1 |
| Inferior lateral ventricle | (Aguilar et al., 2013) | 1 |
| Isthmus cingulate gyrus | (Oliveira Jr et al., 2010) | 1 |
| Orbitofrontal cortex | (Fan et al., 2008b; Wee et al., 2012) | 2 |
| White Matter | | |
| Fornix (WM) | (Dyrba et al., 2015; Westman et al., 2011) | 2 |
| Temporal lobes (WM) | (Davatzikos et al., 2011; Westman et al., 2011) | 2 |
| Ventral cingulum (WM) | (Dyrba et al., 2015) | 1 |
| Caudate nucleus | (Dyrba et al., 2015; Sørensen et al., 2017) | 2 |
| Corpus callosum (WM) | (Dyrba et al., 2015) | 1 |
| Periventricular WM | (Davatzikos et al., 2011; Sørensen et al., 2017) | 2 |
| Parietal WM | (Westman et al., 2011) | 1 |
| Frontal WM | (Westman et al., 2011) | 1 |
| Occipital WM | (Westman et al., 2011) | 1 |
| Inferior temporal WM | (Fan et al., 2008a) | 1 |

## 4.2. Autism

Autism spectrum disorders (ASD) are a series of developmental brain disorders defined by impairment in social interaction, verbal and non-verbal communication and repetitive behavior (Lewis et al., 2013). A few of the works reviewed here use the autism diagnostic observation schedule (ADOS) as a continuous clinical score instead of a binary label (autistic/control). Table 3 shows a summary of the papers reviewed in this section.

Table 3: Summary of the classification papers in autism. Unless otherwise noted, reported accuracy rates are the highest found in the paper for different groups, methods and input modalities.

| Ref | Groups (N) | Method | Input Modalities | Accuracy | Comments |
|---|---|---|---|---|---|
| (Zhou et al., 2014) | Controls (153)-Autism (127) | Multiple | T1, rs-fMRI | 0.7 | Leakage: features selected on the full dataset. Uses 67 different classifiers from the WEKA toolbox. |
| (Ecker et al., 2010a) | Controls (20)-Autism (20) | SVM (linear) | T1 | 0.9 | No hyperparameter search (fixed C = 1). |
| (Sato et al., 2013) | Controls (84)-Autism (82) | SVR (Gaussian) | T1 | r = 0.362 | Predict ADOS scores instead of clinical state as a binary class problem. No hyperparameter search (fixed $\gamma$). |
| (Uddin et al., 2011) | Controls (24) - Autism (24) | SVM (Gaussian) | T1 | 0.92 | Analysis is done per individual region |
| (Libero et al., 2015) | Controls (18) - Autism (19) | Decision tree | T1, DTI, spectroscopy | 0.919 | Possible leakage: *[...] data points included were the significant resulting values of the statistical analyses of separate neuroimaging modalities* |
| (Ingalhalikar et al., 2014) | Controls (42)-Autism (93) ASD/LI+ (36)-ASD/LI- (57) | LDA ensemble | MEG, DTI | 0.83 0.7 | Final accuracy rates are the result of ensembling LDA classifiers that use different combinations of input data. |
| (Wee et al., 2014) | Controls (59)-Autism (58) | SVM (multi-kernel) | T1 | 0.963 | |
| (Ecker et al., 2010b) | Controls (22)-Autism (22) | SVM (linear) | T1 | 0.81 | No hyperparameter search (fixed C = 1). |
| (Lange et al., 2010) | Control (30+7)-Autism (30+12) | QDA | DTI | 0.916 | Independent test set |

Zhou et al. used T1 metrics and network measurements from functional connectivity (rs-fMRI studies) and achieved an accuracy of 0.70. The methodology suffers from leakage (the features have been extracted using the full dataset and not in the cross-validation loop) (Zhou et al., 2014). Ecker et al. report accuracy rates of up to 0.9 when using CT metrics for the left hemisphere. This accuracy drops to 0.6 for the right hemisphere (Ecker et al., 2010a). This significant lateralization is also seen in (Sato et al., 2013), where CT values in the left hemisphere are better predictors of ADOS scores than those of the right hemisphere ($r_{Left}$ = 0.29 vs. $r_{Right}$ = 0.072, $r_{Both}$ = 0.362). A similar effect is also reported in (Uddin et al., 2011), where analyses were made per individual region. In another related work by the same group, patients with higher ADOS scores were found to be further from the optimal hyperplane when a linear-kernel SVM was used for binary classification (Ecker et al., 2010a).

As ASD is a heterogeneous disorder, it has been attempted to fine-tune the definition of the labels to include some of the most relevant symptoms, such as language impairment (ADS/LI+) (Ingalhalikar et al., 2014). However, similar to the case of MCI classification in AD, this task is much more challenging. They obtained an accuracy rate of 0.83 for ASD vs. controls, and 0.7 for ASD/LI vs. ASD/LI+. Additionally, they use model ensemble methods to compensate for missing data.

We have not included a table of relevant regions for this classification problem, since these effects seem to be very broadly spread through the brain in ASD. In addition to the lateralization effect, Wee et al. found that GM values in subcortical regions achieve higher accuracies than cortical regions (Wee et al., 2014).

## 4.3. Multiple sclerosis

As with AD, there is a distinction between healthy controls, clinically isolated syndrome (CIS) (Miller et al., 2012) and fully developed multiple sclerosis (MS). Similarly, classification tasks involving CIS are more challenging. Furthermore, CIS patients have a certain probability of developing MS within a given time window, which is another element to consider. Few papers (summarized in Table 4) have used structural differences for classification in MS. Instead, ML applications have been more heavily focused on the automatic segmentation of WM lesions. This is probably due to the fact that MS diagnosis can be easily made by detecting WM lesions directly from images, and the automatic labeling of those regions is the most challenging part of the problem (García-Lorenzo et al., 2013; Lladó et al., 2012).

Table 4: Summary of the classification papers in MS. Unless otherwise noted, reported accuracy rates are the highest found in each paper for different groups, method and input modalities.

| Reference | Groups (N) | Method | Input Modalities | Accuracy | Comments |
|---|---|---|---|---|---|
| (Wottschel et al., 2015) | CIS (74) - longitudinal (1 year)<br>CIS (74) - longitudinal (3 years) | SVM (polynomial) | T2, PD, clinical, demographic | 0.714<br>0.680 | |
| (Bendfeldt et al., 2012) | Early MS (17) - late MS (17)<br>Low lesion load MS (20)-High lesion load MS (20)<br>Benign MS (13)-Non-benign MS (13) | SVM (linear) | T1, T2 | 0.85<br>0.83<br>0.77 | |
| (Weygandt et al., 2011) | Controls (26) – MS (41) | SVM (linear) | T1, T2 | 0.96 | |
| (Weygandt et al., 2015) | Controls (15+15)-EOPMS (15+16)<br>Controls (15+15)-LOPMS (16+ 17)<br>EOPMS (15+16)-LOPMS (16+17) | Logistic regression | T2 | 0.867*<br>0.871*<br>0.807* | Each voxel individually tested. 2 groups of subjects matched differently (lesion load, gender, & disease duration or age). |

Weygandt et al. used T1 and T2 images to segment the brain into three different regions (lesions and normal-appearing GM and WM) and obtained accuracy rates of up to 0.96 when using lesion information, but also of 0.84 and 0.91 when using normal-appearing regions (GM and WM, respectively) (Weygandt et al., 2011). In a later work, they also obtained high accuracy rates (0.87) when classifying healthy controls vs. early and late-onset pediatric MS (Weygandt et al., 2015). The classification accuracy when comparing the two MS groups was lower (0.807).

Bendfeldt et al. explored classifiers that distinguish between MS subgroups (early or late MS, low WM-lesion load or high WM-lesion load, and benign or non-benign MS) using T1 and T2 data as inputs for linear SVMs. They obtained accuracy rates of 0.85, 0.83, and 0.77, respectively, using GM information alone (Bendfeldt et al., 2012).

Wottschel et al. used 74 subjects at onset of CIS to predict which subjects would develop MS at 1 and 3 years using lesion metrics (count, load, intensity, ...), imaging data and clinical and demographic features (Wottschel et al., 2015). Their results (accuracy scores of 0.714 and 0.68 for 1 and 3 years, respectively), show that the further the time horizon, the harder the classification problem. Also, the optimal feature combinations at 1 year (lesion load, type of presentation, gender) very completely different from the optimal features for the 3-year prediction task (lesion count, average lesion intensity on PD images, average distance of lesions from the center of the brain, shortest horizontal distance of a lesion from the vertical axis, age and Expanded Disability Status Scale (EDSS) at onset).

In terms of region importance, middle frontal gyrus was the most informative in Weygandt et al. (Weygandt et al., 2015), whereas Bendfeldt et al. (Bendfeldt et al., 2012) found relevant regions in cortical areas of all the cerebral lobes, as well as thalamus and caudate.

## 4.4. Parkinson's disease and related disorders

As in some of the previous cases, what initially looks like a binary problem can be further complicated by the introduction of intermediate states or other conditions that are commonly mistaken with the principal disease or disorder. In the case of idiopathic Parkinson's Disease (IPD, or PD), Progressive Supranuclear Palsy (PSP) and Multiple System Atrophy (MSA) have similar motor symptoms, but they also progress faster and are less responsive to treatment (Filippone et al., 2012; Salvatore et al., 2014). Collectively, these are referred to as *Parkinsonian disorders* or *Parkinsonian Plus Syndromes* (Duchesne et al., 2009). A Parkinsonian (MSA-P) and a cerebellar variant of MSA (MSA-C) are distinguished based on clinical presentations (Schulz et al., 1994; Wenning et al., 1994). Recently, another group referred to as SWEDD (Scans Without Evidence of Dopaminergic Deficit) has been added. SWEDD subjects show PD symptoms, but without any dopamine deficiency in their PET scan. Classification tasks in PD therefore include all these disorders as well as the possible combinations including healthy controls. Here we also review papers that use a multiclass approach (i.e. instead of binary classifications, more than two different labels are learned simultaneously) (Filippone et al., 2012; Marquand et al., 2013). Table 5 shows a summary of the relevant findings.

Table 5: Summary of the classification papers in PD.

| Ref | Groups (N) | Method | Input Modalities | Accuracy | Comments |
|---|---|---|---|---|---|
| (Focke et al., 2011) | Controls (22) - PD (21)<br>Controls (22) - PSP (10)<br>Controls (22) - MSA (11)<br>MSA (11) - PD (21)<br>MSA (11) - PSP (10)<br>PD (21) - PSP (10) | SVM (linear) | T1 | 0.42<br>0.937<br>0.788<br>0.719<br>0.762<br>0.968 | Default C hyperparameter (C=1). F-contrast computed using the whole sample applied as weight. |
| (Cherubini et al., 2014) | PD (57) - PSP (21) | SVM (kernel not specified) | T1, T2, DTI | 1 | F-contrast computed using the whole sample applied as weight. |
| (Skidmore et al., 2015) | Controls (22) - PD (20) | Bootstrap | DTI | 0.901 | |
| (Marquand et al., 2013) | PSP (17), PD (14), MSA (19)<br>Controls (19), PSP (17), PD (14), MSA (19)<br>PSP (17), PD (14), MSA-C (7), MSA-P (12)<br>Controls (19), PSP (17), PD (14), MSA-C (7), MSA-P (12) | Multinomial logit | T1 | 0.917<br>0.736<br>0.845<br>0.662 | |
| (Filippone et al., 2012) | Controls (14), PD (14), PSP (16), MSA (18) | Multinomial logit | T1, T2, DTI | Brier= 0.753 | Highest multiclass error score (Brier) obtained using GM only. |
| (Salvatore et al., 2014) | Controls (28) - PD (28)<br>Controls (28) - PSP (28)<br>PD (28) - PSP (28) | SVM (linear kernel) | T1 | 0.927<br>0.970<br>0.982 | Not mentioned how hyperparameters were tuned. |
| (Duchesne et al., 2009) | PD (16) - PSP (8) + MSA (8) | SVM | T1 | 0.906 | PCA transformation applied on 149 healthy controls. No mention on the type of kernel or how hyperparameters were tuned. |
| (Haller et al., 2012) | PD (17) - Other (23) | SVM (Gaussian kernel) | DTI | 0.975 | Heterogeneous "Other" containing patients with different diseases, including MSA and PSP. |
| (Haller et al., 2013) | PD (16) - Other (20) | SVM (Gaussian kernel) | SWI | 0.869 | Same considerations as for (Haller et al., 2012). |

Focke et al. obtained high accuracy rates for controls vs. PSP and PD vs. PSP classifications by using WM voxel values (processed with SPM) as input features (Focke et al., 2011). GM values yielded much lower accuracies. Similarly, in Cherubini et al., WM values alone achieved a perfect classification score (accuracy = 1) (Cherubini et al., 2014). However, important to note that in both cases, F-contrast values were applied as weights for the input voxels out of the cross-validation loop. This could be considered leakage, as this importance metric was computed using the whole sample. Also, the reported WM areas were mainly in the brainstem, where the GM appears as small nuclei surrounded by WM (e.g. *substantia nigra pars compacta*). When using VBM smoothing kernels, these nuclei can appear inside the WM probabilistic mask since the WM signal includes information from both WM and these nuclei.

Both Filippone et al. (Filippone et al., 2012) and Marquand et al. (Marquand et al., 2013) directly build multiclass classifiers. Filippone et al. applied a multinomial logit classifier to a cohort of 62 subjects (14 healthy controls, 14 PDs, 16 PSPs, 18 MSAs) (Filippone et al., 2012). Marquand et al., from the same research group, applied it to a different population and with two variations: a) either healthy controls were included or not in the given classifiers; and b) the MSA cohort was further divided into MSA-P and MSA-C (Marquand et al., 2013). Including healthy controls in the multiclass environment lowered the overall accuracy scores (Marquand et al., 2013). Focke et al. attribute this to inconsistencies in VBM processing (Focke et al., 2011).

In summary, the reviewed results imply PD, PSP and MSA affect different brain regions, even if their symptoms are similar. Relevant regions are summarized on Table 6.

Table 6: Informative regions (GM and WM) for PD classification tasks.

| Region | References | Number |
|---|---|---|
| Grey Matter | | |
| Rectal gyrus | (Skidmore et al., 2015) | 1 |
| Middle cingulate | (Skidmore et al., 2015) | 1 |
| Left Putamen | (Skidmore et al., 2015) | 1 |
| Right Putamen | (Skidmore et al., 2015) | 1 |
| Thalamus | (Haller et al., 2013; Salvatore et al., 2014; Skidmore et al., 2015) | 3 |
| Pons | (Salvatore et al., 2014) | 1 |
| Midbrain | (Marquand et al., 2013; Salvatore et al., 2014) | 2 |
| Brainstem | (Filippone et al., 2012; Marquand et al., 2013) | 2 |
| Caudate | (Filippone et al., 2012; Haller et al., 2013) | 2 |
| Putamen | (Filippone et al., 2012) | 1 |
| Precuneus | (Focke et al., 2011) | 1 |
| Basal ganglia | (Marquand et al., 2013) | 1 |
| Cerebellum | (Marquand et al., 2013) | 1 |
| White Matter | | |
| Corpus callosum | (Salvatore et al., 2014) | 1 |
| Brainstem | (Cherubini et al., 2014; Focke et al., 2011) | 2 |
| Mesoencephalon | (Focke et al., 2011) | 1 |
| Right frontal WM | (Haller et al., 2012) | 1 |

## 4.5. Other

Here we have included diseases or disorders for which we have not found a high number of publications, or in some cases those for which monographic reviews have been published recently.

### 4.5.1. Attention deficit hyperactivity disorder

Iannaccone et al. used both functional and structural imaging to study differences in a cohort of 20 attention deficit hyperactivity disorder (ADHD) patients and 20 healthy controls (Iannaccone et al., 2015). Using only T1 data processed with SPM and a linear SVM (fixed C = 1) they did not obtain a statistically significant accuracy rate (0.611). Lim et al. also used GM information from T1 images (processed with SPM) and a Gaussian process classifier (GPC) and obtained an accuracy of 0.793 for a cohort of 29 ADHD patients and 29 healthy controls (Lim et al., 2013). Finally, Peng et al. achieved up to 0.902 accuracy rates using extreme learning (a neural network variant) using cortical features from T1 data (thickness, surface, folding, curvature, volume) in a cohort of 55 ADHD subjects and 55 healthy controls. However, their feature selection was performed outside of the cross-validation loop (Peng et al., 2013). See also Eloyan et al. for a similar work on same dataset (Eloyan et al., 2012) (for more information, see Section 6 in this paper).

### 4.5.2. Depression

Johnston et al. studied 20 subjects with treatment-refractory depression (TRD) and 21 healthy controls (Johnston et al., 2015). A binary SVM (Gaussian kernel) classifier was able to obtain accuracy rates of 0.85 using T1 images as input. However, it was not possible to produce predictive systems for the level of resistance to treatment. Foland-Ross et al. also used GM information (CT) to separate healthy adolescent girls (n = 15) from those who suffered an initial onset of depression (n = 18) within a 5-year window using linear SVM and obtained an accuracy of 0.7 (Foland-Ross et al., 2015).

As for WM information, using DTI studies, Qin et al. studied network architecture from 29 depressive patients and 30 healthy controls (Qin et al., 2014). Nodal strength, local clustering coefficient, nodal betweenness centrality and nodal global efficiency, for nodes defined in the AAL atlas, were used as input features. Maximum relevance features selection (mRMR) was used to select relevant features in the whole sample (leakage). Under these conditions, a Gaussian SVM obtained a highest accuracy of 0.831. Using a similar approach, Sacchet et al. also used graph theory-related features (assortativity, global flow coefficient, global total flow, global efficiency, characteristic path length, transitivity and small-worldness) as inputs for a linear SVM to distinguish between 14 women with major depressive disorder and 18 healthy controls, obtaining an accuracy of 0.712 (Sacchet et al., 2015).

### 4.5.3. Schizophrenia

For schizophrenia, we refer to recently published reviews that analyze the use of ML algorithms in the context of this disorder in detail. Similar to AD, there are 3 prediction problems of interest in the context of schizophrenia: i) classifying schizophrenia patients versus healthy controls, ii) diagnosing schizophrenia in populations at high risk from baseline scan information, iii) prediction of disease progression, transition to schizophrenia, or response to treatment. Zarogianni et al. provide an extensive review on predictive classifiers for schizophrenia based on either structural or functional MRI, not only focusing on binary predictions, but also devoting a section to disease progression and treatment response (Zarogianni et al., 2013). They report accuracies in the range of 81-91.8% for classifying schizophrenia patients versus healthy controls using sMRI, with the majority of the studies using SVMs for classification. For diagnosing schizophrenia, the reviewed studies have used fMRI as well as sMRI, initially using ICA for dimensionality reduction and mostly SVM and Random Forests for classification, reporting accuracies in the range of 61.8-95%. Fewer studies have attempted to predict transition to schizophrenia and response to treatment, with one study reporting an accuracy of 85% in classifying responders using EEG data and a kernel partial least squares regression technique, and three studies reporting accuracies of 82-84.2% in differentiating transition to schizophrenia, all using SVMs. They conclude that the higher classification accuracy in the first problem (i.e. diagnosing schizophrenia versus healthy controls) is due to the more distinct differences in their neuroanatomical and functional patterns, which is not the case in within group predictions in subjects that do or do not show an specific outcome of interest (Zarogianni et al., 2013). In a more general review, Dazzan also includes a small section on how to use brain structure at illness onset to produce predictions at the individual level (Dazzan, 2014).

### 4.5.4. Traumatic brain injury

We found two studies that build predictive models for traumatic brain injury (TBI), both (Fagerholm et al., 2015; Lui et al., 2014) employing mRMR for feature selection on the entire sample prior to any cross-validation loop (leakage). Lui et al. used T1, DTI and rs-fMRI data for 23 TBI patients and 25 healthy controls, and tested several different classifying algorithms; they obtained an accuracy of 0.86 with a multilayer perceptron (neural network) using only relevant variables, and 0.80 with a Bayesian network using all variables (Lui et al., 2014). Fagerholm et al. used only DTI information, obtaining 24 different graph metrics and an accuracy of 0.934 with a linear SVM (Fagerholm et al., 2015).

### 4.5.5. Stroke

In the context of stroke, machine learning has been used to classify stroke patients versus normal controls, or predict post-stroke functional impairment or treatment outcome. Rehme et al. used DTI and resting state fMRI data information and a linear SVM to classify stroke patients vs. normal controls (accuracy= 0.826), and predict motor impairment after stroke (accuracy=0.876). They also used information from DWI lesion maps to differentiate stroke patients with or without hand motor impairment, but with a relatively low sensitivity (accuracy= 0.738, sensitivity= 0.50), concluding that resting state fMRI is more useful in predicting behavioural deficits than DTI (Rehme et al., 2014). Bently et al. used CT information in combination with clinical variables and an SVM with a multi-layer perceptron kernel to predict whether or not to administer thrombolysis, a treatment that can result in better recovery or deterioration due to intracranial haemorrhage (AUC=0.744) (Bentley et al., 2014).

### 4.5.6. Miscellanea

*Anorexia nervosa (AN).* Lavagnino et al. used a LASSO regression to classify 15 patients with AN and 15 healthy controls using T1 information (processed with FreeSurfer), obtaining a accuracy of 0.833 (Lavagnino et al., 2015).

*Bipolar disorder (BD).* Hajek et al. obtained an accuracy of 0.689 with a linear SVM (fixed C = 1) when differentiating 45 healthy subjects from 45 high-risk offsprings from subjects with BD (Hajek et al., 2015). Only using WM intensities (T1 scans, processed with SPM) yielded significant accuracies. Similarly, 36 healthy controls were distinguished from 36 BD patients with an accuracy of 0.597. In all experiments, subjects were matched by age and sex.

Lastly, we reference Sabuncu and Konukoglu, an empirical review that applies several ML algorithms to different data sets for a variety of diseases and disorders in a standardized way to build a *gold standard* that can be used to compare the accuracy of future new approaches (Sabuncu et al., 2015).

## 5. Discussion

In this review, we compile an extensive summary on ML techniques in the field of neuroimaging, from cross-validation analyses to specific applications in different diseases or disorders using structural modalities. We have attempted to include a wide-reaching sample that will help the reader get a precise grasp of the current state of the art. ML applications in this field are different from applications in other areas such as spam (or credit card fraud) detection. In the clinical case, practical application of ML does not simply aim to achieve the highest accuracy scores possible, as is the case when filtering spam e-mails, for instance. While it is undoubtedly preferable to obtain higher accuracy rates, in neuroscience, it is more relevant to study which features are informative for the classification task of interest as well as their corresponding biological interpretations (see for instance (Carbonell et al., 2015)).

A common pattern seen in the literature is that classification tasks are almost never *purely* binary in nature. While two-class approximations are still relevant (AD vs. controls, MS vs. controls, PD vs. controls, etc.), including intermediate (MCI, CIS, etc.) or related (PSP, MSA, etc.) states can add another level of complexity. In practical terms, the common solution is to opt for multiple binary comparisons (AD vs. controls, controls vs. MCI, AD vs. MCI), each of which can be solved by developing a separate classifier whose performance is assessed individually. Only in a few cases (e.g. (Filippone et al., 2012; Marquand et al., 2013)) has a multiclass approach been used. While binary approaches provide useful information about underlying biological mechanisms, from a clinical point of view, multiclass approaches might be more insightful, as a binary classifier would require to eliminate *a priori* all potential clinical labels but two, a process that is not always practical. This is further complicated by the fact that many disorders are *spectrum* disorders, and therefore a binary variable may not completely capture their underlying subtleties.

The best classification performances were obtained when differentiating between normal controls versus patients in various diseases (e.g. AD, autism, MS, PD) with accuracies higher than 0.9, suggesting the existence of brain patterns and structures identifiable on MRI that are significantly different between the diseased population and normal controls and can be reliably used for differentiating these groups (Table 7). Unfortunately, the accuracies were much lower (generally around 0.7) when attempting to differentiate between progressive and stable patients (e.g. MCI-c and MCI-nc in AD), although these problems are of higher clinical interest. While the reviewed studies provide valuable benchmarks for classification accuracy, in practice, there's still a need for double-blind experiments. Clinical trials in which the predictions are made before the actual outcome (e.g. conversion to AD) has been observed can provide confirmatory evidence for the clinical use of the prognosis models. Additionally, challenges that are administered by a different research group and provide only the necessary MRI and clinical data without the outcomes of interest on a preserved test dataset (e.g. MICCAI conference TADPOLE challenge: https://tadpole.grand-challenge.org/) would also ensure that the results are not influenced by leakage or overfitting the models.

Table 7 compares the results of the studies that classify normal controls versus patients. While generalizations made based on such small sample sizes and numbers of studies should be taken into consideration with care, the number of the studies in each field seem to reflect the current view on the structural nature of diseases (as can be detected on MRIs). The neurodegeneration pattern that is characteristic of AD seems to be a very good indicative for clinical diagnosis. On the other hand, most studies that attempt to make diagnosis for Schizophrenia use function modalities which might hint at the insufficiency of structural MRI for such predictions. Another factor that needs to be considered is the very different sample sizes across studies in different diseases, e.g. AD studies generally have much larger sample sizes. While this is influenced by the disease prevalence as well as financial funding allocations, the amount of available data on AD can considerably facilitate studies in this field.

Table 7: Summary of the studies differentiating between normal control and patients.

| Disease | Methods | Input Modalities | Accuracy Mean | Accuracy Min - Max | Number of Studies |
|---|---|---|---|---|---|
| Alzheimer's Disease | SVM, OPLS, Random Forests | T1, PET, DTI, CSF | 0.897 | 0.82 – 0.965 | 19 |
| Autism | SVM, Decision Tree, LDA, QDA | T1, DTI, Spectroscopy | 0.867 | 0.70 – 0.963 | 8 |
| Multiple Sclerosis | SVM, Logistic Regression | T1, T2 | 0.915 | 0.871 – 0.96 | 2 |
| Parkinson's Disease | SVM, Bootstrap, Multinomial Logit | T1, T2, DTI | 0.7472 | 0.42 – 0.927 | 5 |
| Attention Deficit Hyperactivity Disorder | SVM, Gaussian Process Classifier | T1 | 0.8475 | 0.793 – 0.902 | 2 |
| Depression | SVM | T1, DTI | 0.7477 | 0.70 – 0.831 | 3 |

Sometimes, certain processing techniques can be tuned for a specific disease or disorder in order to take into account some aspect that could improve the overall accuracy. Take for example Dukart et al., who removed age effects when noticing an age difference in misclassified individuals depending on the cohort (AD or healthy subjects) (Dukart et al., 2011).

We acknowledge there are limitations to the present study. First, this review focuses on analyses that employ only structural MRI data (T1, T2, and DWI), while we have also included works that have used other imaging modalities, either in isolation or in combination with structural MRI. However, in practice, other imaging modalities as well as a battery of clinical tests and measurements are acquired which can provide informative features that might significantly improve the classifications. For example, in the case of converter versus non-converter MCI subjects in the context of AD prediction, using the baseline clinical information significantly improves the prediction accuracy (Moradi et al., 2015). Since different studies acquire different MRI modalities and clinical information, we were not able to compare them across all modalities and measures. However, we have reported the other modalities and measures that have been used (e.g. fMRI, PET, clinical measures, etc.) for each study. Additionally, it has to be noted that in some cases (e.g. (Dyrba et al., 2015)), the inclusion of additional modalities does not increase accuracy. More features can result in more information, but also in more noise and confounding factors (Duda et al., 2001).

It should be taken into account that the output labels (the clinical state for each subject) may be an approximation, as there may not be a precise one-to-one correspondence between a given metric (e.g. ADOS) and a binary clinical outcome. Also, the clinical diagnosis might contain errors (Cherubini et al., 2014), and therefore it would impossible to obtain a perfect classification.

In some cases, diseases or disorders encapsulate a gradient of symptoms and causes. Additionally, the outcome of interest might be the amount or rate of change in a given metric (e.g. cognitive or motor function) rather than simply whether the subject declines or not. Such problems would be better studied using continuous regression techniques and not discrete classifications. A number of regression techniques have been used for estimating continuous clinical variables using neuroimaging data, such as linear regression and support vector regression (Duchesne et al., 2009, 2005; Hope et al., 2013; Rondina et al., 2016; Stonnington et al., 2010; Wang et al., 2010; Zhang et al., 2011). However, this review focuses on predictions that can be formulated as a binary classification task.

Another point that is worthwhile mentioning is the outcome of interest, which can be different for different prediction problems and in different populations in the clinical setting. For example, positive predictive value (the percentage of correct positive predictions over all positive predictions) or negative predictive value (the percentage of correct negative predictions over all negative predictions) might have more clinical relevance in specific cases. For the purpose of consistency and since it is the most commonly used measure across papers, here we report classification accuracy which reflects the percentage of both negative and positive correct predictions over all predictions.

While there are hundreds of different ML algorithms (Fernández-Delgado et al., 2014), there is undoubtedly a preponderance of SVMs in the neuroimaging literature (Table 7) (Liu et al., 2012). This goes so far as some reviews (e.g. (Salvatore et al., 2014; Veronese et al., 2013)) are centered exclusively on using SVMs for predictive purposes. While this can be attributed to the fact that previous experience greatly influences the choice of a certain algorithm, it is also true that SVMs behave robustly in the typical conditions of a neuroimaging problem: i.e. many more variables than available subjects (in some cases, these differences are of several orders of magnitude) (Lemm et al., 2011; Zarogianni et al., 2013). It has to be also taken into account, however, that other techniques have also been employed with comparable results (see Table 1, for instance) and that feature selection techniques, when correctly applied to avoid leakage, are extremely useful in reducing the dimensionality of the problem.

A fraction of the works included in this review report potentially overly-optimistic results (leakage), due to the fact that informative variables were selected outside of the cross-validation loop. This feature selection step was performed using statistical techniques that assess group differences in the entire sample, for instance, or used other types of filtering procedures that relied on the class labels of the test set to perform dimensionality reduction. As discussed before (Section 3.2.1), this should be avoided, as it might produce biased results.

Leakage effect is especially important when the population size is small and diminishes as the sample size grows. Kohavi and John reported this effect to be less concerning when the dataset contains more than 250 instances (Kohavi and John, 1997). However, this number is also dependent on the choice of classifier and the number of features used. Almost all the studies reviewed here have sample sizes smaller than 250, which makes the leakage issue more prominent. It would be interesting to compare different studies in the same domain to assess whether the reported accuracies are significantly different in cases where leakage occurs. However, since different studies are based on different populations and features, drawing meaningful comparisons is not feasible in practice.

Another challenge that can reduce the generalizability of classification models to new data and consequently their applicability to clinical practice is the inherent heterogeneity in neuroimaging datasets. Imaging data from different scanner and acquisition protocols can sometimes have very different contrasts and parameters. As a result, the estimated performances and classifier accuracies may only be reliable when applied to data from similar scanners and with similar acquisition parameters as the training dataset. Several preprocessing pipelines have been developed to deal with such variabilities, such as the SPM, FSL, and MINC tools (Aubert-Broche et al., 2013; Jenkinson et al., 2012; Penny et al., 2011). In addition, to increase the generalizability of the results, models are generally trained on multi-site and multi-scanner datasets (such as ADNI, PPMI, etc.).

Throughout this article, we have reviewed papers typically written in research institutions by domain experts: either scientists that have a close contact with clinical environments or more technically-oriented individuals who, also in a clinical or biomedical context, find in these datasets the opportunity to apply and improve their current algorithms. In recent years, non-domain experts (pure ML engineers, mathematicians, etc.) have also had access to datasets already processed, and have attempted to solve these classification challenges. From this point of view, websites such as Kaggle (https://www.kaggle.com) do a great job in gathering ML experts around a very heterogeneous set of problems. This has been the case, for instance, with the IEEE International Workshop on Machine Learning for Signal Processing (MLSP) 2014 Schizophrenia Classification Challenge (https://www.kaggle.com/c/mlsp-2014-mri), the American Epilepsy Society Seizure Prediction Challenge (https://www.kaggle.com/c/seizure-prediction), or the Predict HIV progression Challenge (https://www.kaggle.com/c/hivprogression). These challenges typically work in the following way: datasets are provided for both a training set and a test set. The output labels (clinical classifications) are also provided for the training set, and the competitors have to produce their predictions for the test set, with any technique they wish; the only common restriction is to use technologies with open-source licenses. Submissions are evaluated according to a certain metric (for instance, the AUC score) and the different teams are ranked in a classification table (*leaderboard*). To prevent competitors from training their predictors to simply increase the final scores in the leaderboard, these scores are computed on an unknown portion of the test set. These competitions normally take several months to complete and some have associated monetary prizes for the highest ranked teams.

Iannaccone et al. present a similar application in their Introduction: The ADHD-200 Global Competition (http://fcon_1000.projects.nitrc.org/indi/adhd200/results.html) (Iannaccone et al., 2015). In this challenge, functional and structural imaging and demographic and behavioral data was provided with the aim of producing individual clinical predictions for ADHD subjects. Interestingly, the highest accuracy (0.625) was obtained by one of the competing teams using only age, sex, handedness, and IQ, and no imaging information. Other recent examples include the Medical Image Computing and Computer Assisted Intervention (MICCAI) 2014 Machine Learning Challenge: Predicting Binary and Continuous Phenotypes from Structural Brain MRI Data (https://competitions.codalab.org/competitions/1471) and its sister challenge CADDementia (Bron et al., 2015).

ML algorithms are powerful tools that can be used to solve many different problems. In the field of neuroscience, these powerful tools can not only help us build predictive systems for diagnosis and prognosis, but can also be used to advance and deepen our knowledge about the underlying biological mechanisms of diseases and disorders.

# Acknowledgments

Authors have received funding from the Brain Canada Foundation and *Fondation Marcelle et Jean Coutu*. Authors also wish to thank Dr. Penelope Kostopoulos for her insightful comments.